\documentstyle[twoside,fleqn,espcrc2,epsf]{article}

\newcommand{\AmS}{{\protect\the\textfont2
  A\kern-.1667em\lower.5ex\hbox{M}\kern-.125emS}}

\title{Computation of the Heavy-Light Decay Constant
       with NRQCD}

\author{Shoji Hashimoto\address{Department of Physics, Hiroshima University,
                                Higashi-Hiroshima 724, Japan}
}

\begin{document}

\begin{abstract}
Non-relativistic QCD is applied for a lattice computation
of the heavy-light meson decay constant in quenched approximation at
$\beta=6.0$.
Clear signals are obtained for the ground state at large times in
the correlators, allowing a reliable extraction of the decay constant.
Estimating the current renormalization factor by the tadpole
improvement procedure, we find $f_{B}=$ 164(17) MeV for the $B$
meson with $a^{-1}=$ 2.3 GeV, while an extrapolation to the static limit
yields $f_{B}^{\mbox{static}}=$ 247(26) MeV.
\end{abstract}

\maketitle

\section{Introduction}

In recent years large effort has been directed toward a lattice QCD
calculation of the decay constant of heavy-light mesons.  Uncertainties
in the results, however, are still quite large\cite{Bernard}.  In the static
approximation for heavy quarks, the problem manifests in  the presence of
large
noise in the correlators, which makes it difficult to extract ground state
signals.  Smearing techniques have not resolved the problem, failing to yield
results independent of the smearing size\cite{Hashimoto_Saeki}.

In this work we report on a calculation of heavy-light decay constants using
non-relativistic QCD (NRQCD)\cite{Thacker_Lepage}.  Our motivation
stems from the expectation that noise in
correlators should be reduced for a finite heavy quark mass\cite{Lepage} and
hence cleaner results could be obtained in NRQCD.  In addition NRQCD allows
incorporation of $1/m_Q$ corrections in a systematic way.
This is an important point since  available results indicate that
$1/m_{Q}$ corrections to the static limit is fairly large
for the $B$ meson\cite{Bernard}.

\section{Simulation}

For the heavy quark we employ the standard NRQCD action given
by\cite{Thacker_Lepage}
\begin{equation}
S_{Q}^{(n)}= a^{3} \sum_{x t} Q_{x t}^{\dagger} \left[\Delta_{4}- H^{(n)}
             - c_{1} \frac{\vec{\sigma}\cdot\vec{B}}{2 m_{Q} a}
            \right] Q_{x t}
\end{equation}
where $H^{(1)}=H=\sum_j\Delta_{-j}\Delta_j/2m_Qa$ and $H^{(2)}=H-H^2/4$, the
latter being a modified choice in order to stabilize high frequency modes for
$m_Qa<3$. The spin-magnetic interaction term is included for keeping
consistency of the $1/m_{Q}$ expansion.
We use the tree-level value $c_1=1$  for the coefficient.

Our simulation is carried out with  40 quenched configurations on an
$\mbox{16}^{3} \times 48$ lattice at $\beta = \mbox{6.0}$.
The heavy quark masses used are $m_Qa=1000, 10.0, 7.0, 5.0, 4.0$
with the $n=1$ action and $5.0, 4.0, 3.0, 2.5$ with $n=2$.
The value $m_{Q}=1000$ is taken to  compare with the results of the static
approximation.  For the light quark we used the Wilson action with
$K=\mbox{0.1530}$, $\mbox{0.1540}$ and $\mbox{0.1550}$.
The critical hopping parameter is $K_{c}=0.15708(2)$.

The heavy-light meson decay constant is extracted from the correlator of
the axial-vector current and the heavy-light meson in the standard manner.
For the meson operator we use the local form and also employ
the cube smearing of sizes
$\mbox{3}^{3}$, $\mbox{5}^{3}$, $\mbox{7}^{3}$ and $\mbox{9}^{3}$,
fixing gauge configurations to the Coulomb gauge
to avoid the necessity of inserting gauge link factors.

\section{Results}

\subsection{Signal to noise ratio}

\begin{figure}
\epsfxsize85mm
\epsfbox{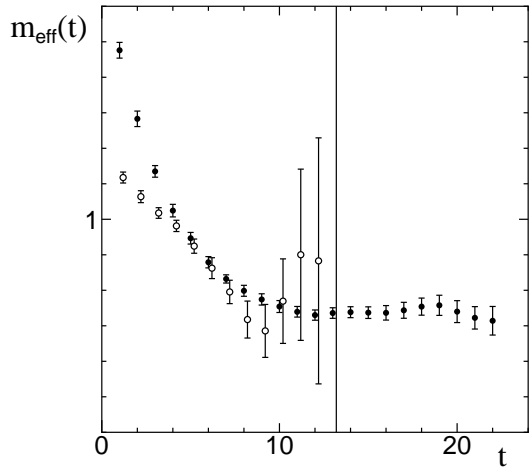}
\caption{Effective mass of the local-local correlation function for
$m_Qa=1000$
         (open circles) and $m_Qa=5$ (filled circles),
          both with $K=0.1530$ for light quark.}
\label{figone}
\end{figure}

In Fig.~\ref{figone} we show the effective mass of the local-local correlation
function
\begin{equation}
C(t)=<\bar{Q}(t)\gamma_4\gamma_5 q(t) \bar{q}(0)\gamma_5Q(0)>
\end{equation}
for
$m_{Q}a$=$1000$ and 5.0.  Clearly the signal is far better for
$m_{Q}a$=5.0 for which we observe a plateau beyond $t\approx 10-12$.

\begin{figure}
\epsfxsize85mm
\epsfbox{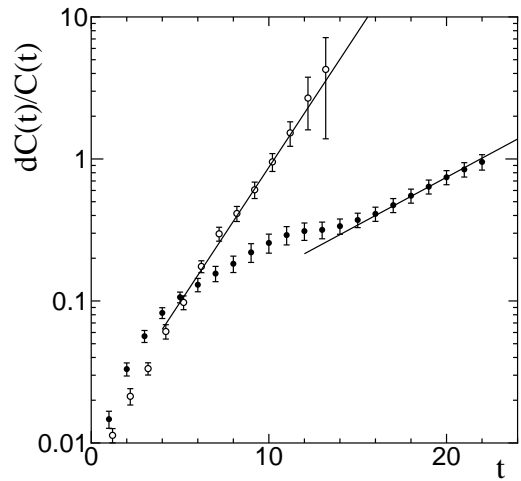}
\caption{Relative error of local-local correlation function for $m_Qa=1000$
         (open circles) and $m_Qa=5$ (filled circles),
          both with $K=0.1530$ for light quark.}
\label{figtwo}
\end{figure}

The improvement  of ground state signals for smaller values of $m_Qa$ can be
qualitatively understood from the estimate of the relative error\cite{Lepage},
\begin{equation}
\frac{\delta C(t)}{C(t)} \propto
       \exp \left[ \big(E(Q\bar{q}) - \frac{E(Q\bar{Q}) + m_{\pi}}{2}\big) t
\right] \end{equation}
\noindent
where $E(Q\bar{q})$ and $E(Q\bar{Q})$ are binding energies
of heavy-light and heavy-heavy mesons.
For finite $m_Qa$ the negative contribution of $E(Q\bar{Q})$
reduces the value of the exponential slope
from that in the static limit where
$E(Q\bar{Q})$ vanishes.
We found that our data for $\delta C(t)/C(t)$ are
consistent with this estimate.
Typical examples are shown in Fig.~\ref{figtwo}
where solid lines indicate the slope expected from the
measured values of the binding energies and $m_\pi$.

\subsection{Dependence on smearing size}

\begin{figure}[htb]
\epsfxsize85mm
\epsfbox{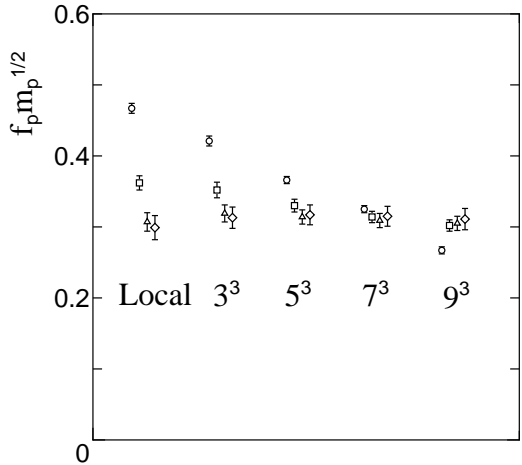}
\caption{Dependence of \protect$f_P\protect\sqrt{m_P}\protect$
        on the fitting range
        $t_{min}\leq t\leq t_{min}+4$ for various smearing sizes
        at $m_Qa=$ 5.0 and $K=$ 0.1530.}
\label{figthree}
\end{figure}

In Fig.~\ref{figthree} we plot the raw value of the combination
$f_P\sqrt{m_P}$ of
the decay constant $f_P$ and the heavy-light meson mass $m_P$ extracted from
fits of the correlators over the interval $t_{min}\leq t\leq t_{min}+4$ for
various smearing sizes.
For each group of data points $t_{min}$ increases as
$t_{min}=$ 6, 8, 10, 12 from left to right.
We observe that the estimates converge to the same value after $t\approx 10$
for all the smearing sizes including the case of no smearing.
This gives us confidence that the asymptotic region is really reached
at $t\approx 10-12$.
We also note that the magnitude of errors
are similar for various smearing sizes.  In particular reliable results can be
obtained without smearing in NRQCD.
In the
following analysis we use the cube smearing of size $\mbox{5}^{3}$ and extract
$f_P\sqrt{m_P}$ from a global fit over the interval $10 \leq t \leq 20$.
Other choices give similar results.

\subsection{Heavy-light decay constant}

\begin{figure}[htb]
\epsfxsize85mm
\epsfbox{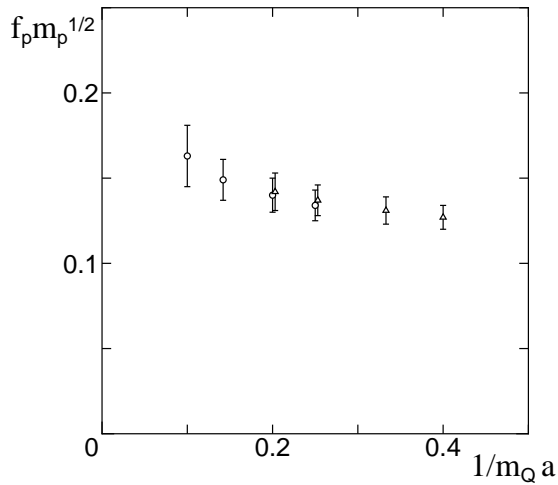}
\caption{ $1/m_{Q}$ dependence of
          \protect$f_{P} \protect\sqrt{m_{P}}\protect$ after extrapolation the
limit $K=K_c$ for light quark.}
\label{figfour}
\end{figure}

In order to obtain $f_{P} \sqrt{m_{P}}$ as a function of the  heavy
quark mass $m_{Q}a$ we extrapolate the results at three values of $K$ for the
light quark linearly in $1/K$ to $K_c$ for each $m_Q$.
For the axial-vector current renormalization factor $Z_A$
we take the value $Z_A=0.65$
obtained by applying the improvement procedure with the coupling
$g^2_V(1/a)$\cite{LM} to the one-loop result\cite{Davies_Thacker_2},
disregarding a small $m_Q$ dependence of a few \% over our range of $m_Qa$.
The results for $f_{P} \sqrt{m_{P}}$ are plotted in Fig.~\ref{figfour}.
Circles and triangles are for the  results obtained with the
$n=1$ and $2$ action in (1).  We see that they yield consistent values for
$f_{P} \sqrt{m_{P}}$ in the region of $m_Qa$ where both  actions can be
employed.

In order to see the deviation from the heavy mass scaling law
$f_{P} \sqrt{m_{P}}=\mbox{const}$,
we fit this data with the form,
\begin{equation}
f_{P} \sqrt{m_{P}} = (f_{P} \sqrt{m_{P}})^{\mbox{static}}
                     ( 1 - \frac{c}{m_{Q} a} ),
\label{fit}
\end{equation}
and obtain
\begin{equation}
(f_{P} \sqrt{m_{P}})^{\mbox{static}}=\mbox{0.163(17)},
\quad c=\mbox{0.60(20)}.
\end{equation}
With $a^{-1}=$ 2.3(1) GeV obtained from the $\rho$ meson mass
on the same set of configurations, this gives
\begin{equation}
f_B^{\mbox{static}}=247(26)
\left(\frac{a^{-1}}{\mbox{2.3 GeV}}\right)^{3/2} \mbox{MeV}
\end{equation}
for the $B$ meson decay constant in the static approximation.

Substituting $m_Qa=1.8$ for the $b$ quark mass
which is estimated from the $\Upsilon$ mass
including the heavy quark mass renormalization\cite{Davies_Thacker_1},
we find a substantially smaller value for the $B$  meson decay
constant in NRQCD:
\begin{equation}
f_B=\mbox{164(17)} \left(\frac{a^{-1}}{2.3\mbox{GeV}}\right)^{3/2}
\mbox{MeV}.
\end{equation}
The large coefficient $c$=0.60(20) in (\ref{fit}) leads to a significant
$1/m_{Q}$ correction ($\sim$ 30 \%) to the static approximation, which  is
consistent with the results
obtained with propagating heavy quarks\cite{Bernard}.

\section*{Acknowledgements}
The author thanks M.~Okawa and A.~Ukawa for valuable discussions.
Numerical calculations were made on HITAC S820/80 at KEK.
This work is supported in part by the Grant-in-Aid of
the Ministry of Education (No. 040011).


\begin{thebibliography}{9}
\bibitem{Bernard}See, {e.g.,} review by C. Bernard, in these proceedings.
\bibitem{Hashimoto_Saeki}
   S. Hashimoto and Y. Saeki, Mod. Phys. Lett. {\bf A7} (1992) 387;
   C. Bernard {\it et al.}, {\it Lattice 91}, Nucl. Phys. B(Proc. Suppl.)
   {\bf 26} (1992) 384.
\bibitem{Thacker_Lepage}
   G. P. Lepage and B. A. Thacker, Nucl. Phys. B(Proc. Suppl.) {\bf 4}
   (1988) 199;
   B.A. Thacker and G.P. Lepage, Phys. Rev. {\bf D43} (1991) 196.
\bibitem{Lepage}
   G.P. Lepage, Nucl. Phys. B (Proc. Suppl.) {\bf 26} (1992) 45.
\bibitem{LM} G.~P.~Lepage and P.~ Mackenzie, Phys. Rev. {\bf D48} (1993) 2250.
\bibitem{Davies_Thacker_2}
   C.T.H. Davies and B.A. Thacker, Phys. Rev {\bf D45} (1992) 915;
   Phys. Rev {\bf D48} (1993) 1329.
\bibitem{Davies_Thacker_1}
   C.T.H. Davies and B.A. Thacker, Nucl. Phys. {\bf B405} (1993) 593.


\end{thebibliography}
\end{document}